\documentclass[journal]{IEEEtran}
\usepackage{cite}
\usepackage{amsmath,amssymb,amsfonts}
\usepackage{algorithmic}
\usepackage{graphicx}
\usepackage{textcomp}
\usepackage{gensymb}
\def\BibTeX{{\rm B\kern-.05em{\sc i\kern-.025em b}\kern-.08em
    T\kern-.1667em\lower.7ex\hbox{E}\kern-.125emX}}
\newcommand{\mycircle}[1]{\negthinspace
\raisebox{.5pt}{\normalsize\textcircled{\raisebox{-.5pt}
{\scriptsize#1}}}\negthinspace
}
\makeatletter
\def\ps@IEEEtitlepagestyle{%
  \def\@oddfoot{\mycopyrightnotice}%
  \def\@evenfoot{}%
}
\def\mycopyrightnotice{%
  {\footnotesize \copyright~20XX IEEE\hfill}
  \gdef\mycopyrightnotice{}
}

\begin{document}
\title{Physics-Coupled Neural Network Magnetic Resonance Electrical Property Tomography (MREPT) for Conductivity Reconstruction} 
\author {
Adan Jafet Garcia Inda, \IEEEmembership{Member, IEEE},
Shao Ying Huang, \IEEEmembership{Member, IEEE},
Nevrez \.{I}mamo\u{g}lu, \IEEEmembership{Member, IEEE}, and
Wenwei Yu, \IEEEmembership{Member, IEEE}%
\thanks{Adan Jafet Garcia Inda is with the Department of Medical Engineering, Chiba University, Chiba, Japan. (e-mail: adanjgi@gmail.com)}%
\thanks{Shao Ying Huang is with the Department of Surgery, National University of Singapore, Singapore, Singapore and the Engineering Product Development Department, Singapore University of Technology and Design, Singapore, Singapore. (e-mail: huangshaoying@sutd.edu.sg)}%
\thanks{Nevrez \.{I}mamo\u{g}lu is with the Digital Architecture Research Center, National Institute of Advanced Industrial Science and Technology, Tokyo, Japan. (e-mail: nevrez.imamoglu@aist.go.jp)}%
\thanks{Wenwei Yu is with the Center for Frontier Medical Engineering, Chiba University, Chiba, Japan and the Department of Medical Engineering, Chiba University, Chiba, Japan. (e-mail: yuwill@faculty.chiba-u.jp)}}%

\markboth{IEEE TRANSACTIONS ON IMAGE PROCESSING, ~Vol.~XX, No.~XX, SEPTEMBER XX~2021}%
{Shell \MakeLowercase{\textit{et al.}}: Bare Demo of IEEEtran.cls for IEEE Journals}

\maketitle
\begin{abstract}
The electrical property (EP) of human tissues is a quantitative biomarker that facilitates early diagnosis of cancerous tissues. 
Magnetic resonance electrical properties tomography (MREPT) is an imaging modality that reconstructs EPs by the radio-frequency field in an MRI system.
MREPT reconstructs EPs by solving analytic models numerically based on Maxwell’s equations. 
Most MREPT methods suffer from artifacts caused by inaccuracy of the hypotheses behind the models, and/or numerical errors. 
These artifacts can be mitigated by adding coefficients to stabilize the models, however, the selection of such coefficient has been empirical, which limit its medical application. 
Alternatively, end-to-end Neural networks-based MREPT (NN-MREPT) learns to reconstruct the EPs from training samples, circumventing Maxwell’s equations.
However, due to its pattern-matching nature, it is difficult for NN-MREPT to produce accurate reconstructions for new samples.
In this work, we proposed a physics-coupled NN for MREPT (PCNN-MREPT), in which an analytic model, cr-MREPT, works with diffusion and convection coefficients, learned by NNs from the difference between the reconstructed and ground-truth EPs to reduce artifacts. 
With two simulated datasets, three generalization experiments in which test samples deviate gradually from the training samples, and one noise-robustness experiment were conducted. 
The results show that the proposed PCNN-MREPT achieves higher accuracy than two representative analytic methods. Moreover, compared with an end-to-end NN-MREPT, the proposed method attained higher accuracy in two critical generalization tests.
This is an important step to practical MREPT medical diagnoses.
\end{abstract}


\IEEEpeerreviewmaketitle

\section{Introduction}
\label{sec:introduction}
\IEEEPARstart{M}{agnetic} resonance imaging (MRI) visualizes human tissues by using the interaction of magnetization, radiofrequency (RF) waves, and protons in the tissue. MRI shows good soft-tissue contrast. 
However, for early stages cancers, until the cancerous tissues are calcified, they are out of the reach of T$_1$- or T$_2$-weighted MRI\,\cite{Wu2009}.
On the other hand, Electrical properties (EPs) are quantitative bio-markers that help discern various pathologies and their development\,\cite{Jensen-Kondering2020}. 
Images using EPs as tissue contrasts may offer a way for early detection of cancers\,\cite{Bodenstein2009,Seo2014}. 
Furthermore, spatial EPs distribution is an inevitable base to calculate specific absorption rate (SAR) for the assessment of RF safety in any environment that exposes tissues to RF/microwave waves\,\cite{Wolf2013}, e.g., wireless power transfer in a human involved environment\,\cite{zhou2021_WPT}. 

Magnetic resonance electrical properties tomography (MREPT) uses the measured RF fields in an MRI scanner, the $B_1$ fields, to reconstruct the EPs of the tissues of the human body under scan\,\cite{Haacke1991}.
To calculate/reconstruct EPs in MREPT, the $B_1$ maps (magnitudes and the phase) are required\,\cite{Katscher2009}, for which the magnitudes of $B_1^+$ and $B_1^-$ and the phase sum of $B_1^+$ and $B_1^-$, ($\phi^+$\,+\,$\phi^-$), can be measured from an MRI scanner\,\cite{Sacolick2010,Cunningham2006}. 
The reconstruction of EPs can be realized either by solving an analytic formulation of Maxwell's equations numerically or by data-driven end-to-end methods, which obtain a function mapping $B_1^+$ fields to EPs directly from training samples, neglecting Maxwell's equations.

Assumptions underlie most analytic MREPT methods.
One of these is the homogeneity assumption which presupposes null spatial changes of the EPs to simplify the formulation\,\cite{Wen2003}. It is a straightforward approach, however, it produces artifacts at the tissue boundaries which are ubiquitous in the human body\,\cite{huang2015_FEM_MREPT,Mandija2018}.
Another assumption is that the conductivity has a close relation to the phase sum ($\phi^+$\,+\,$\phi^-$), and the permittivity has a close relation to the $B_1^+$ magnitude\,\cite{Hoult2000}. 
This assumption decouples the reconstructions for conductivity and permittivity to produce independent reconstruction formulations that shorten MRI's scan time. 
While these assumptions facilitate the application of the MREPT method to clinical fields\,\cite{Tha2018,VanLier2014}, they result in and further amplify artifacts in the reconstructions\,\cite{Mandija2018,Shin2014}.
Various methods have been developed to address the artifact problems of MREPT caused by the above-mentioned assumptions. 
For example, to actualize the homogeneity assumption, T$_1$- or T$_2$-weighted MRI image-based segmentation is used to produce EP reconstructions for each homogeneous segment\,\cite{Katscher2012}, which relies on the assumption that EP segmentation from T$_1$- or T$_2$-weighted MRI images are accurate. 
Other proposed methods require $B_1$\,-\,field measurements that are impossible in current MRI scanners\,\cite{Nachman2007}, or involves multiple MRI scans to solve the reconstruction problem\,\cite{Zhang2010}, which increases scan time, an undesired effect in the clinical field. 

A proposed method in\,\cite{Hafalir2014} removes the homogeneity assumption and offers flexibility to solve the reconstruction problem.
In\,\cite{Hafalir2014}, the reconstruction was formulated as a convection reaction partial differential equation (cr-EPT), then discretized and solved for the region of interest (ROI). 
By taking this approach, EPs are unknowns to solve without any spatial constraint, thus the homogeneity assumption is no longer needed.
However, this method results in numerical artifacts when the resolution of the $B_1$ maps is low\,\cite{Ider2020}. 
The effect of numerical artifacts in cr-EPT can be reduced by two methods, either by increasing the discretization density or by adding an artificial diffusion term, $\boldsymbol\rho\nabla^2\gamma$, where $\boldsymbol\rho$ is called the diffusion coefficient\,\cite{Li2017}.
The former increases computation cost exponentially, thus it is not preferable. 
In the latter, the added term plays the role of viscosity regularization that reduces sharp variations in the reconstructed EPs, including numerical artifacts, without increasing the computation cost. However, with a large $\boldsymbol\rho$, this term dominates the equation, which blurs the EP distribution, lowering the contrast. 
Therefore, this term requires an appropriate diffusion coefficient to achieve a trade-off between the reduction of the sharp variations in the EP reconstruction and a decrease in the contrast. In\,\cite{Li2017}, $\boldsymbol\rho$ is suggested to be no more than 1 for good EPs reconstructions. 
In\,\cite{Li2017,Gurler2017,Sun2020}, this coefficient has been selected empirically and applied homogeneously to the ROI, i.e., a single global coefficient for the whole ROI.
\begin{figure*}[!ht]
	\centerline{\includegraphics[width = 0.9 \textwidth]{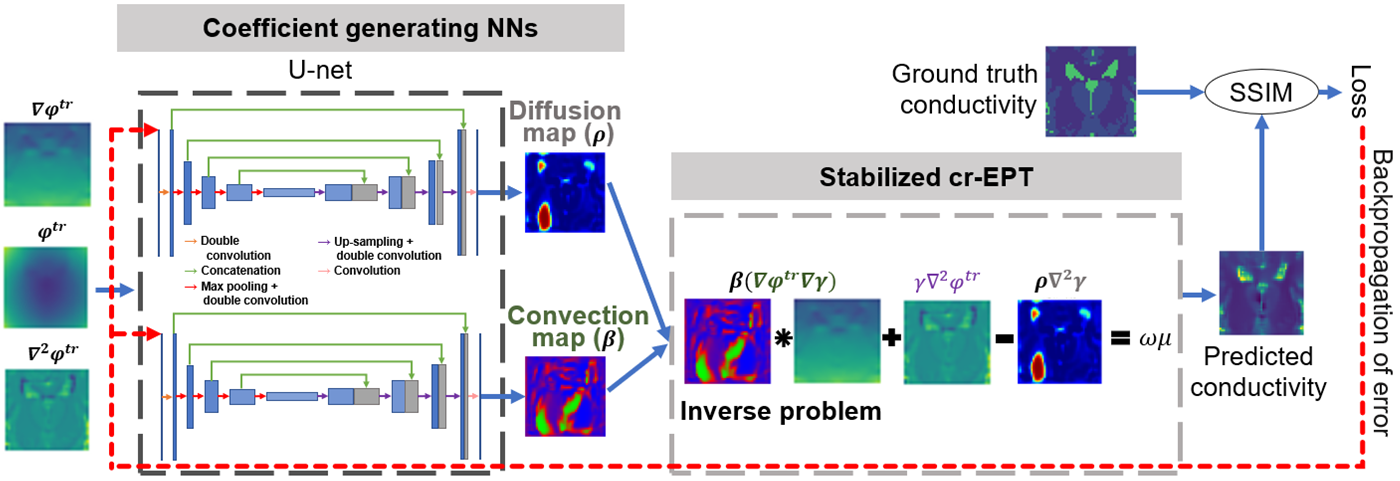}}
	\caption{The proposed physics-coupled-neural-network electrical properties tomography (PCNN-EPT) that consists of the coefficient generating NNs (U-net) and the stabilized cr-EPT. The former assigns random diffusion coefficient ($\boldsymbol\rho$) maps and convection coefficient ($\boldsymbol\beta$) maps to produce a conductivity reconstruction by the stabilized cr-EPT. 
	Structural similarity index measure (SSIM) between the ground truth and the predicted conductivity is used as a loss function to feedback to the NNs to optimize the reconstruction coefficients maps, to increase the accuracy of the conductivity reconstruction.}
	\label{fig:Diagram}
\end{figure*}

In our previous work\,\cite{Garcia2018}, a pixel-wise diffusion coefficient (called local diffusion map) was found to increase the reconstruction contrast better than a global diffusion coefficient for the whole ROI. 
The reason behind the advantage of the local diffusion map is that the boundary areas where artifacts are more likely to appear in the reconstructed EPs require different diffusion values compared to non-boundary areas for artifacts suppression.
However, the local diffusion map was empirically set depending on the geometric structure of EPs, which is only possible for simple geometries and requires prior knowledge about the EPs to be reconstructed. 
The difficulty of setting the local diffusion map lies in that for every pixel its coefficient stabilizes the spatial variations of not only the designated pixel but also the surrounding ones due to the effects of the diffusion term.
Effective stabilization of spatial variations in a pixel-wise manner for an accurate EP reconstruction is a pixel-wise interconnected high dimensional problem. Thus, optimization without knowing the ground truth of the local diffusion maps for human tissues of complex geometries, for unknown samples is necessary for the medical application of MREPT.

On the other hand, in cr-EPT, besides the diffusion coefficient ($\boldsymbol\rho$), a global artificial convection coefficient, ($\boldsymbol\beta$), was recently suggested to adjust the weight of the convection term to correct boundary artifacts\,\cite{Ider2020}. In\,\cite{Ider2020}, a single $\boldsymbol\beta$ was proposed for the whole ROI rather than pixel dependent. 
For $\boldsymbol\beta$, different regions (boundaries or non-boundaries) in the ROI may need different values to improve the quality of reconstruction. Again, optimization of $\boldsymbol\beta$ in a pixel-wise manner, without knowing the ground truth of it is necessary. 
Meanwhile, different roles of $\boldsymbol\beta$ need to be identified and analyzed for tuning this coefficient to benefit the reconstruction. 
Furthermore, the interactions and cooperation between $\boldsymbol\rho$ and $\boldsymbol\beta$, globally or locally, for EP reconstruction may offer a much higher degree of freedom to improve the reconstruction accuracy, which remains to be explored.

Alternatively, to avoid the artifacts associated with the numerical solution of analytic MREPT, data-driven end-to-end methods have been explored for EPs reconstruction. 
MREPT reconstructions have been achieved by using training samples to guide neural networks (NNs) to learn a function mapping $B_1^+$ fields to EPs\,\cite{Mandija2019,Garcia2020,Garcia2019}. 
These end-to-end NN-based methods do not capture the wave physics contained in analytic MREPT, thus they can be categorized as physics-unaware approaches.
Due to this lack of wave physics, the physics-unaware NN-based approach requires huge amounts of data to ensure its robustness to new samples, i.e., generalization. However, the collection of huge amounts of data is costly, especially in the medical field.
Moreover, it was further observed that the NN-based MREPT reconstructions were susceptible to various types of noise\,\cite{Hampe2020} that are common in MRI. 
In general, noise robustness can be expected from NNs, due to its training sample interpolation nature, but only if the training dataset contains samples contaminated by appropriate types of noise.
However, it is impossible to obtain “ideal” noisy samples from a clinical MRI system because the ground truth EPs distribution of real tissues could be neither measured non-invasively nor obtained through accurate reconstruction by any current MREPT methods. 
Moreover, although electromagnetic simulations could provide samples with ground truth EPs, the lack of exact noise models for clinical MRI systems makes it difficult to generate realistic noisy training samples. 
Therefore, it is of extreme importance to explore methods that acquire reconstruction models inherently robust to different types of noise, likely to achieve high generality from fewer training samples (i.e., high sample efficiency).

Recently, physics-aware NN-models have shown sample efficiency improvements\,\cite{Bar-Sinai2019,Raissi2019,Aggarwal2019}. In these data-driven physics-aware approaches, physics knowledge is embedded into NNs models derived from corresponding mathematical formulations.
In these works, NNs were trained 1) to replace a mathematical operation, such as first-order or second-order derivative, which is likely subjected to noise or numerical computation error\,\cite{Raissi2019,Bar-Sinai2019}; 2) to emulate the input-output relationship of a physics phenomenon\,\cite{Carleo2017}; 3) to play a role of model-based filtering, as a pre-processing or a post-processing operation\,\cite{Aggarwal2019}.
In 1) and 2), the data-driven process is separated from the physics models, thus physics models only influence the learning through the training dataset, under the assumption that physics models do not have any variant, e.g., models with different parameters, or activation modes.
In 3) Physics models work as an active part to feedback modeling error to drive the adaptation process of the filtering function, though the data-driven process does not alternate any parameters of the physics models themselves. 
In summary, in all the current physics-aware approaches, physics-awareness is incomplete in terms that physics models are assumed to be rigid (fixed), and either NNs or physics models affect the other in a unidirectional way, thus they could not generate a solution that can deal with a diversity of samples for high generality.
As aforementioned, MREPT has well-established physics models with enriched wave physics and mathematical formulations, though it requires problem-specific local diffusion and convection parameters for solving the boundary artifact problem and stabilize the reconstruction of EPs with geometric and distributional complexity. 
Moreover, there has been neither empirical knowledge nor ground truth about the two physics parameters. Therefore, a data-driven approach needs to be coupled with analytic models for acquiring the model parameters from EPs samples, while seeking reconstruction with high accuracy and high generality.

In this work, we proposed a physics and NNs coupling framework: physics-coupled neural network electrical properties tomography (PCNN-EPT), in which a model derived from Maxwell equations solves the MREPT reconstruction problem with the help of two physics-aware regularization mechanisms. These mechanisms are optimized by two NNs updated with the backpropagated gradient computed from the errors of the EPs reconstructed by the analytic model.
The framework was implemented with cr-EPT as the analytic model, in which the regularization mechanisms are the two stabilization coefficients introduced above, $\boldsymbol{\rho}$ and $\boldsymbol{\beta}$, optimized by NNs.
This proposed approach offers high flexibility for the cooperation of $\boldsymbol{\rho}$ and $\boldsymbol{\beta}$ region by region to achieve the best reconstruction accuracy. 
Furthermore, it can handle the high dimensionality of the problem and is explainable because the NNs produced physics-aware regularization mechanisms on the analytic model offer an analytic foundation rather than a black-box model. 
To make clear the role of the two coefficients, the local $\boldsymbol{\rho}$ and local $\boldsymbol{\beta}$ are compared with its counterparts, namely, the different combination of local and global (scalar or single-value coefficient) of $\boldsymbol{\rho}$ and $\boldsymbol{\beta}$. 

\section{Formulation of stabilized cr-ETP}
\label{sec:formulation}
Phase-based MREPT\,\cite{Voigt2011} is a fast approach due to the reduced MRI scan time\,\cite{VanLier2012}.
The simplest form, as shown in\,\eqref{eqn:ph-std-EPT}, corresponds to the standard EPT formulation (std-EPT)\,\cite{Wen2003} where homogeneous conductivity in the ROI is assumed, i.e., $\nabla\gamma = 0$. 
Moreover, it is assumed that the transmit and receive phases are similar, and the transmit phase could be approximated as half of the transceive phase. 
This assumption is known as the "half phase assumption"\,\cite{Wen2003}.
In 3 Tesla or lower conditions, this assumption's impact is not very significant\,\cite{VanLier2014}.
\begin{equation}
\label{eqn:ph-std-EPT}
    \nabla^2 \phi^{\text{tr}} \gamma=\omega\mu_0  
\end{equation}
where $\phi^{\text{tr}}$ is the transmit phase from the $B_1^+$ field using the transceive phase assumption\,\cite{VanLier2012}, $\omega$ is the Larmor frequency, $\gamma$ is the inverse of the conductivity $\gamma\negmedspace=\negmedspace 1/{\sigma}$. 
Due to the assumption on the homogeneity of the conductivity, the reconstructed conductivity by \eqref{eqn:ph-std-EPT} results in artifacts near the boundaries.
These boundary artifacts can be dampened by solving the spatial changes of the conductivity maps as unknowns.
The convection reaction-EPT (cr-EPT) formulation as shown in \eqref{eqn:ph-cr-EPT} is the approach in this direction derived from Maxwell's equations\,\cite{Gurler2017}.
\begin{equation}
    (\nabla\phi^{\text{tr}} \nabla\gamma)+\gamma\nabla^2 \phi^{\text{tr}} =\omega\mu_0  
    \label{eqn:ph-cr-EPT}
\end{equation}

Although the assumption on the homogeneity of the conductivity is removed by cr-EPT, it presents numerical artifacts in the solution due to the numerical instability associated with the discretization.
The cr-EPT model could be complemented with an artificial diffusion term ($\boldsymbol{\rho}\nabla^2\gamma$) to reduce the numerical artifacts\,\cite{Li2017}, and an artificial convection coefficient ($\boldsymbol\beta$) to cooperate with the diffusion term to produce accurate contrast\,\cite{Ider2020}, leading to the equation below, 
\begin{equation}
    \boldsymbol\beta(\nabla\phi^{\text{tr}} \nabla\gamma)+\gamma\nabla^2 \phi^{\text{tr}} -\boldsymbol\rho\nabla^2 \gamma=\omega\mu_0  
    \label{eqn:ph-double-stab-EPT}
\end{equation}
$\boldsymbol{\rho}$ is called diffusion coefficient. In the literature, constant $\boldsymbol{\rho}$'s are empirically selected to the whole ROI\,\cite{Li2017,Gurler2017,Sun2020}, so are $\boldsymbol{\beta}$'s\,\cite{Ider2020}. 

\section{Methods and materials}
\label{sec:methods-materials}
\subsection{The proposed method}
\label{PCNN-EPT}
Fig.\,\ref{fig:Diagram} shows the flow of the proposed PCNN-EPT where the formulation of the physical model is based on the phase-based stabilized cr-EPT\,\cite{Hafalir2014} shown in \eqref{eqn:ph-double-stab-EPT}. 
As shown in Fig.\,\ref{fig:Diagram}, the input of PCNN-EPT includes the normalized transmit phase ($\phi^{\text{tr}}$), its gradient ($\nabla\phi^{\text{tr}}$), and its Laplacian ($\nabla^2\phi^{\text{tr}}$). 
The proposed method consists of two parts, the coefficient generating NNs and the stabilized cr-EPT.
The former is a two-stream NN structure, each of which produces one of the two stabilization coefficients, $\boldsymbol\beta$, and $\boldsymbol\rho$.
The generated $\boldsymbol\beta$ and $\boldsymbol\rho$ are passed to the stabilized cr-EPT to compute the conductivity map as the output.
The predicted conductivity map is compared with the ground-truth conductivity map from the training samples, and a structural similarity index measure (SSIM) based loss function\,\cite{Huang2021} is calculated and back-propagated to the coefficient generating NNs updating their parameters, which is indicated by the red dashed line.

\subsubsection{Coefficient generating Neural networks}
\label{subsubsec:coefficient generating neural-networks}
The coefficient generating NNs learn to produce the stabilization coefficients from its inputs, and the EPs reconstruction errors, without explicit ground truth of the coefficients, which is unknown in the MREPT research area. This ground truth unknown model may bypass the bias brought by training samples. 
A randomly initialized U-net architecture\,\cite{Ronneberger2015} is employed to map the $\phi^{\text{tr}}$\,inputs to the stabilization coefficients, due to U-net's local connectivity in multiple receptive fields. 
Two individual NNs with the same number of parameters are used to produce $\boldsymbol\beta$ and $\boldsymbol\rho$ separately. 
The outputs of the NNs are bounded by a sigmoid activation function at the output to prevent the stabilizing coefficients from overwhelming the reconstruction formulation in \eqref{eqn:ph-double-stab-EPT}. The $\boldsymbol\rho$ is bounded to [1$\times$10$^{\text{-6}}$,\,0.1] while the $\boldsymbol\beta$ coefficient is bounded to [0.01, 5].
Without bounding $\boldsymbol\beta$, the convection term may overtake the reconstruction during learning and the formulation can be overshadowed by this term. 
On the other hand, without bounding $\boldsymbol\rho$, the viscosity term may over-tighten the bonds among the values of the pixel in the neighborhood, dampening all spatial variations.
Equation \eqref{eqn:optim-coeff} below shows the formulation of the optimization for the i$^\text{th}$ iteration.
\begin{equation}
   \mathbf{\rho}_{\theta_{\text i}},\mathbf{\beta}_{\theta_{\text i}} = \operatorname*{argmin}_{\theta \in \Theta} \mathcal{L}[\sigma,  \hat{\sigma}{(\mathbf{\rho}_{\theta_{\text{i}-1}},\mathbf{\beta}_{\theta_{\text{i}-1}})}]
    \label{eqn:optim-coeff}
\end{equation}
where $\mathbf{\rho}_{\theta_{\text{i}}}$ and $\mathbf{\beta}_{\theta_{\text{i}}}$ are the $\boldsymbol\rho$- and $\boldsymbol\beta$- coefficients at the i$^\text{th}$ iteration, given the NN parameters $\theta_i$, which belong to a parameter space ($\Theta$), $\mathcal{L}$ is a loss function that is defined in \eqref{eq:lossfn1} using SSIM\,\cite{Huang2021} in \eqref{eq:lossfn2},
$\hat{\sigma}{(\mathbf{\rho}_{\theta_{\text{i}-1}},\mathbf{\beta}_{\theta_{\text{i}-1}})}$ is the predicted conductivity on the previous iteration, and $\sigma$ is the ground truth.
In \eqref{eqn:optim-coeff}, $\hat{\sigma}{(\mathbf{\rho}_{\theta_{\text{i}-1}},\mathbf{\beta}_{\theta_{\text{i}-1}})}$ is compared to $\sigma$ through the loss function $\mathcal{L}$. 
The NNs parameters $\theta_{\text{i}}$ are optimized at the i$^\text{th}$ training iteration to minimize the loss function $\mathcal{L}$, according to its gradients by an RMSprop\,\cite{Tieleman2012} optimizer algorithm with a starting learning rate of 0.001 and 0.9 momentum.
\begin{equation}
       \mathcal{L_{\text{SSIM}}} = 1-\text{SSIM}(\sigma,\hat{\sigma})
\label{eq:lossfn1}
\end{equation}
\begin{equation}
       \text{SSIM}(\sigma,\hat{\sigma})=\frac{(2\mu_{\sigma}\mu_{\hat{\sigma}} +c_{1})(2\,\text{SD}_{\sigma,\hat{\sigma}}+c_{2})}{(\mu_{\sigma}^2+\mu_{\hat{\sigma}}^2+c_{1})(\text{SD}_{\sigma}^2+\text{SD}_{\hat{\sigma}}^2+c_{2})}
\label{eq:lossfn2}
\end{equation}
where $\sigma,\hat{\sigma}$ are the ground truth and reconstructed conductivity, respectively, $\mu$ represents the mean value of a set patch of pixels, SD represents the standard deviation of the same patch, and $c_1$ and $c_2$ are prefixed coefficients\,\cite{Wang2004}.
SSIM is chosen from different options to construct the loss function, for overcoming extremely local artifacts that appear in the reconstructed conductivity by \eqref{eqn:ph-double-stab-EPT}. This is possible because the $\mu$ and SD in \eqref{eq:lossfn2} calculation is performed over an area i.e., a patch of multiple pixels, which reduces the impact of the local artifacts.
If pixel-wise error measures such as MSE were used, the training will be biased by these local artifacts, resulting in high diffusion values ($\boldsymbol\rho\approx0.1$), which reduce the overall spatial variations and cause EPs reconstruction with very low contrast.

For an independent comparison, a PCNN-EPT model where the stabilization coefficients are scalar values applied globally in the ROI is also estimated (PCNN-global-EPT).
For PCNN-global-EPT, to generate these coefficients, two parallel two-layered fully connected architectures are used. 
On the first layer, the inputs are compressed to the square root of the phase image size and on the second layer, it is further compacted to a scalar output.

Additionally, a physics-unaware (end-to-end) data-driven method was taken for comparison. It is an enforced implementation of\,\cite{Mandija2019}, referred to as NN-EPT.
The loss function of NNs for NN-EPT is based on mean squared error (MSE), because, the NN-EPT could benefit more from MSE than SSIM, due to the nonexistence of local artifacts in the end-to-end method. Moreover, the loss function comparison can be found in Appendix \ref{subsec:NN-EPT}.
\begin{figure*}
    \centerline{\includegraphics[width = 0.9\textwidth]{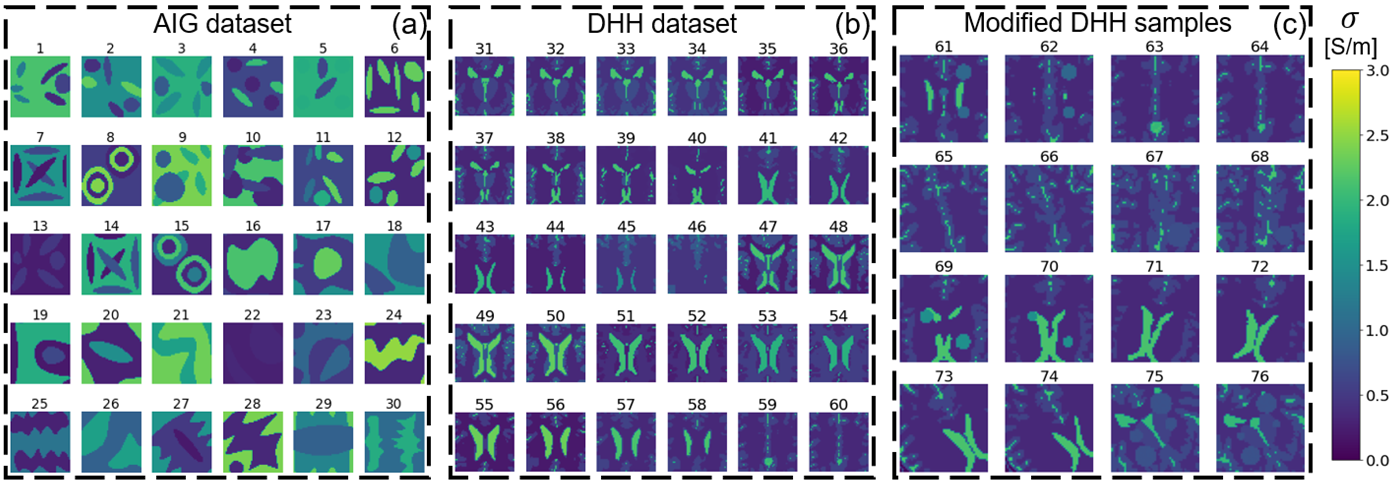}}
    \caption{Datasets, (a) samples \#1-30,  artificial irregular geometries (AIG) dataset, (b) samples \#31-60, digital human head (DHH) dataset (samples \#31-46 obtained  from "Duke" model, samples \#47-60 obtained from "Ella" model), (c) samples \#61-75 digital human head with added variations, the model present added masses and rotations up to 15\degree\,.}	
\label{fig:data}
\end{figure*}
\begin{figure}
    \centerline{\includegraphics[width=0.43\textwidth]{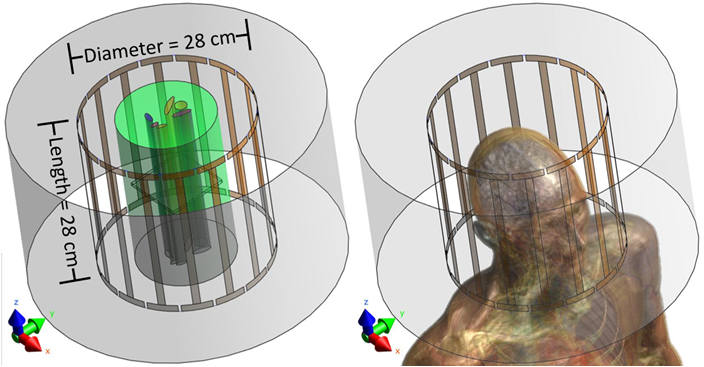}}
	\caption{Birdcage-coil on Sim4Life$^\copyright$ excited at 128\,MHz (3\,Tesla) in quadrature mode, loaded with a cylindrical sample with an artificial irregular geometry (left) and with a digital human head (right).}
    \label{fig:structures}
\end{figure}
\subsubsection{Stabilized cr-EPT}
\label{subsubsec:Analytical stabilized-EPT}
The stabilized cr-EPT is based on the formulation in \eqref{eqn:ph-double-stab-EPT}, taking the outputs of the coefficient generation NNs, local/global $\boldsymbol{\rho}$ and $\boldsymbol{\beta}$, as inputs, and generating a conductivity map as output.
2-D reconstruction is focused in this study for a reduced computational cost, with a potential to be extended to 3-D cases. 
The spatial resolution is 2\,mm in the x-, y- and z-direction.
The central difference of each point of the ROI is taken, by using the finite difference method to produce a mesh to solve the formulation in \eqref{eqn:ph-double-stab-EPT}. 
The phase-based stabilized cr-EPT formulation after discretization is shown in \eqref{eqn:discretized-EPT} below,
\begin{equation}
    \begin{aligned}
        \omega\mu_0 = \sigma_{\text{i,j}}\left[\frac{\delta^{2}\phi_{x}^{\text{tr}}}{\delta x^2}+\frac{\delta^{2}\phi_{y}^{\text{tr}}}{\delta y^2}\right]\\
        +\boldsymbol\beta_{\text{i,j}}\left[\frac{\delta\phi_{x}^{\text{tr}}(\sigma_{\text{i+1,j}}-\sigma_{\text{i-1,j}})}{2\delta x}+\frac{\delta\phi_{y}^{\text{tr}}(\sigma_{\text{i,j+1}}-\sigma_{\text{i,j-1}})}{2\delta y}\right]\\ 
        -\boldsymbol\rho_{\text{i,j}}\left[\frac{\sigma_{\text{i+1,j}}-2\sigma_{\text{i,j}}+\sigma_{\text{i-1,j}}}{\delta x^2}
        +\frac{\sigma_{\text{i,j+1}}-2\sigma_{\text{i,j}}+\sigma_{\text{i,j-1}}}{\delta y^2}\right]
    \end{aligned}
    \label{eqn:discretized-EPT}
\end{equation}
where $i$ and $j$ are the index number for pixels in the x- and the y-direction, respectively, the $\phi^{\text{tr}}$ derivatives are calculated using the $2^\text{nd}$ degree polynomial Savistky-Golay filter. 
With the discretization in \eqref{eqn:discretized-EPT}, a set of linear equations can be obtained for the ROI and form an $\boldsymbol{Ax}=\boldsymbol{b}$ system where $\sigma_{i,j}$ lies in the unknown array, $\boldsymbol{x}$. 

\subsection{Dataset Generation}
\label{subsec:datasets}
The inputs to the EPT algorithms/networks, $\phi^{\text{tr}}$, $\nabla\phi^{\text{tr}}$, $\nabla^2\phi^{\text{tr}}$ were prepared based on simulated $B_1^+$ of different phantoms.
The simulated $B_1^+$ datasets were generated by Sim4Life$^\copyright$ (ZMT AG, Zurich).
Fig.\,\ref{fig:structures} shows the phantoms built. As shown, the transmit coil is a shielded high-pass 16-rung birdcage coil. 
The inner diameter was set to be 28\,cm for head imaging. 
The birdcage coil is excited by two ports separated 90$\degree$ apart geometrically, and excited with a harmonic excitation at 127.78\,MHz and a phase shift of 90$\degree$ between the ports for a quadrature mode. It has a shield with a diameter of 50\,cm.
Complex $B_1^+$ fields associated with the specific slices for reconstruction were extracted, and the transceive phase $\phi^{\text{tr}}$ and the corresponding derivatives were used as inputs for the models. 
The coil is loaded with different phantoms to produce the training samples required for the data-driven models. 
Moreover, complex $B_1^+$ fields at the center slides along the z-direction were selected. 
This reduces the end-ring effects and suppresses the values of $\delta B_z$, thus it is assumed that $\delta B_z\!\approx\!0$.

Two different types of numerically simulated phantoms are produced.
One is a 2-D artificial irregular geometries (AIG) dataset shown in Fig.\,\ref{fig:data}\,(a), the other is a digital human head (DHH) dataset shown in Fig.\,\ref{fig:data}\,(b) generated from the virtual population 3.0\,\cite{Gosselin2014}. 
The AIG dataset is composed of 30 samples produced with a main cylindrical structure of 16\,cm in diameter and 24\,cm in length with various foregrounds embedded into it. The length of the cylinder was set to suppress the end effects. 
The structures vary in size and in conductivity (0.1\,-\,2.5\,S$/$m). 
This dataset was produced to present samples that vary in both conductivity values as well as in major structure, providing diverse information to the reconstruction models.
The second type of dataset from the digital human head model presented in \cite{Gosselin2014} is produced by 15 different models with conductivity changes of $\pm$40\%.
The 15 models correspond to 8 male digital models ("Duke") and 7 female models ("Ella"). 
From each model, two continuous slices from the ROI were picked to be reconstructed.
The heights of the male and female models are different but the ROI is set to a fixed height, this produces different anatomical structures to be focused on for either the male or female models.
This dataset was produced to present samples with similar anatomical structures and complex boundaries as in the clinical field.

\subsection{Generalization}
\label{subsec:gen}
Three experiments were conducted to examine the generalization of the proposed PCNN-EPT. 
The test samples are designed to deviate from the training samples at three levels, from the first experiment with the least deviation to the third one with the most deviation.

In Experiment\,1 (Exp.\,1), the AIG dataset in Fig.\ref{fig:data}\,(a) and the DHH dataset in Fig.\ref{fig:data}\,(b) were divided in an equally divided 5-fold cross validation scheme.
The mean reconstruction accuracy from the test samples from all folds summarizes the accuracy of the learning approaches for the dataset.

In Experiment\,2 (Exp.\,2), the five NN models trained with the dataset from Fig.\ref{fig:data}\,(b), and the test samples are 16 digital head models with modifications like rotations or additional pathological tissues, as shown in Fig.\ref{fig:data}\,(c). 
The added rotations on the models are within 15\degree\,in either direction.
The added pathological models are spheres from 0.8 to 2.4\,cm in diameter at random locations of the brain in the ROI, with an increase of 60-120\% in conductivity of that of the white matter to mimic a brain tumor\,\cite{Hancu2019}. 
The modified test samples represent variations that may appear in clinical practice. 

In Experiment\,3 (Exp.\,3), the five models trained using the DHH dataset in Fig.\,\ref{fig:data}\,(b) are tested with six different test samples from the AIG dataset in Fig.\,\ref{fig:data}\,(a) each. 
Exp.\,3 explores dataset generalization, i.e., the responses of the proposed model on difficult cases for data-driven methods where the samples presented to the network are completely different from those used during training.

\subsection{Noise Robustness}
\label{subsec:noise-method}
Besides the generalization, another experiment (Exp.4) was designed and conducted to evaluate the noise robustness of the proposed PCNN-EPT without any further training.
Using the trained models from the previous experiments, the noise robustness of the methods is tested by adding noise to its corresponding test samples from each experiment. 
The transceive phase $\phi^{\text{tr}}$ was contaminated with additive zero-mean Gaussian distributed noise.
The added noise was calculated by simulating the noise of the $B_1^+$ magnitude image produced by a double angle $B_1^+$ retrieval method\,\cite{Cunningham2006}.
Thus, the relation of the standard deviation (SD) of the noise, $\text{SD}_{\phi^{\text{tr}}}$, and the signal-to-noise-ratio (SNR) of the magnitude image is expressed below\,\cite{Hafalir2014}:
\begin{equation}
    \text{SD}_{\phi^{\text{tr}}}=\frac{\sqrt{2}}{\text{SNR}}
    \label{eq:Noise-SD}
\end{equation}
The noise levels were set to be in the range of those of clinical MRI systems.
The noisy $\phi^{\text{tr}}$ and its derivatives were used as the inputs of the model.
A pre-process low-pass Gaussian filter with a standard deviation of one was applied before the numerical derivative to avoid noise explosions\,\cite{Gurler2017}.
This process was applied equally to all the methods under comparison when noisy samples were tested.

\section{Results}
\label{sec:results}
\begin{figure*}[!ht]
	\centerline{\includegraphics[width=\textwidth]{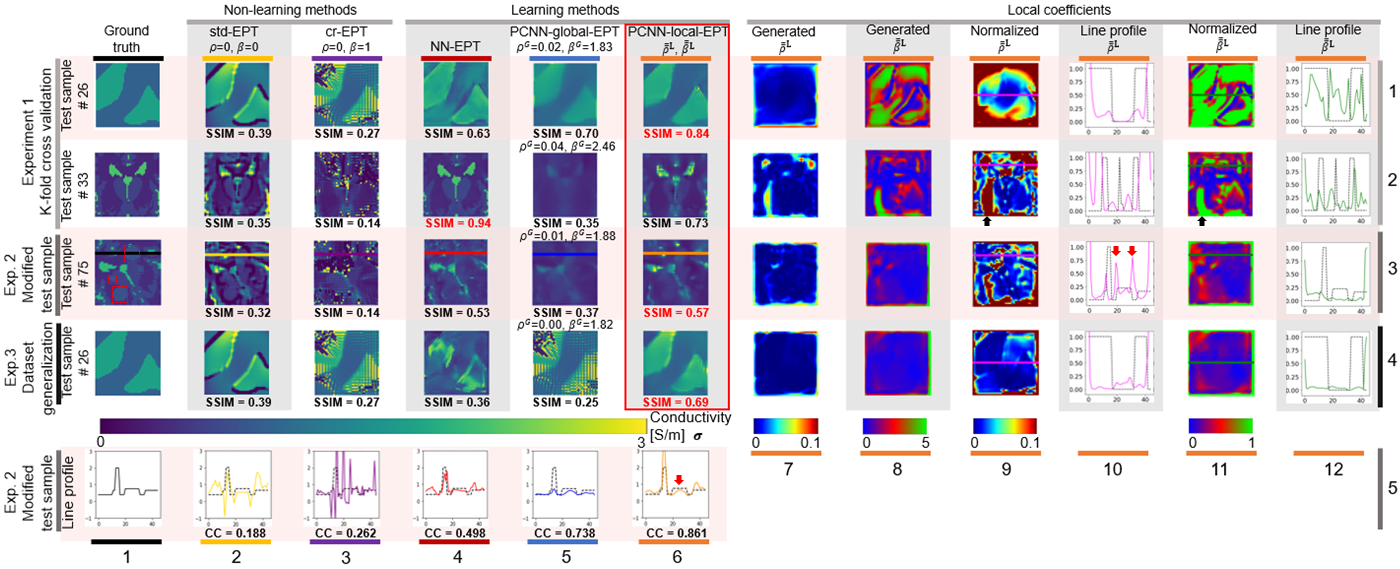}}
	\caption{Reconstructed conductivity maps in Experiment\,1\,-\,3 for generalization by the non-learning numerical models (std-EPT, cr-EPT), and the learning methods (a physics-unaware method (NN-net) and the proposed PCNN-EPTs). The PCNN-EPT includes the PCNN-global-EPT with learned stabilization scalar coefficients $\mathit{\rho}^\textit{G}$ and $\mathit{\beta}^\textit{G}$ and the PCNN-local-EPT with learned stabilization matrix coefficients $\Bar{\Bar{\mathbf{\rho}}}^\textbf{L}$ and $\Bar{\Bar{\mathbf{\beta}}}^\textbf{L}$ at Column\,7 and 8, respectively. SSIM values are shown for each reconstruction. To visualize the coefficients clearly, the normalized  $\Bar{\Bar{\mathbf{\rho}}}^\textbf{L}$ and $\Bar{\Bar{\mathbf{\beta}}}^\textbf{L}$ along with their respective line profiles are shown at Column\,9\,,\,10\,,\,11\,and\,12. The last row shows the line profiles for the sample reconstructions at Row\,3 and their corresponding correlation coefficient (CC) compared to the ground truth. PCNN-local-EPT is the only learning method that maintains its contrast across all rows.}
	\label{fig:Results_compiled}
\end{figure*}
\subsection{Examination of Generalization}
\label{subsec:recon results}
Fig.\,\ref{fig:Results_compiled} shows the reconstructed EP maps for several samples using the proposed PCNN-EPT as well as those from analytic methods (std-EPT, cr-EPT) and the physics-unaware learning method (NN-EPT). The SSIM values of the reconstructed conductivity maps are included.
The conductivity ground truth for each case is included at Column\,1.
The reconstructed conductivity maps by the non-learning methods, std-EPT, and cr-EPT, are shown at Column\,2 and\,3, respectively. 
The physics-unaware learned NN-EPT reconstructions are shown at Column\,4.
For the PCNN-global-EPT, the learned stabilization coefficients are scalar values, $\mathit{\rho}^\textit{G}$, and $\mathit{\beta}^\textit{G}$ are shown on top of each reconstruction at Column\,5, whereas for the PCNN-local-EPT, the learned stabilization coefficients are matrices, $\Bar{\Bar{\mathbf{\rho}}}^\textbf{L}$ and $\Bar{\Bar{\mathbf{\beta}}}^\textbf{L}$, where each pixel has a learned $\boldsymbol\rho$ or $\boldsymbol\beta$. 
The learned $\Bar{\Bar{\mathbf{\rho}}}^\textbf{L}$ and $\Bar{\Bar{\mathbf{\beta}}}^\textbf{L}$ for the PCNN-local-EPT reconstructions in Column\,6 are shown at Column\,7 and 8, respectively. 

Moreover, the learned $\Bar{\Bar{\mathbf{\rho}}}^\textbf{L}$ and $\Bar{\Bar{\mathbf{\beta}}}^\textbf{L}$ are normalized to enhance its visualizability with maximum values of 0.1 and 1, respectively, and presented at Column\,9 and 11. 
Lastly, the line profiles of $\Bar{\Bar{\mathbf{\rho}}}^\textbf{L}$ and $\Bar{\Bar{\mathbf{\beta}}}^\textbf{L}$ with the normalized ground truth conductivity line profile (black dashed line) in the background are shown in their adjacent columns at Column\,10 and 12, respectively. 
It is worth noting that the line profile of $\Bar{\Bar{\mathbf{\rho}}}^\textbf{L}$ is increased five-fold to visualize its spatial shifts.

At Row\,1 and 2, the reconstructed conductivity maps using the learning methods are from Exp.\,1, 5-fold cross-validation, where both test and training samples are from the same dataset, the AIG or the DHH dataset shown in Fig.\,\ref{fig:data}\,(a) and (b), respectively.

The reconstructed EPs based on the learning model at Row\,3 are from the Exp.\,2 where the test samples are from the modified DHH samples in Fig.\,\ref{fig:data}\,(c) and the training samples are from the DHH dataset in Fig.\,\ref{fig:data}\,(b).
At Row\,4, the reconstructed EP maps of the learning methods are for Exp.\,3, the most challenging one, where test and training samples are completely different. The former are from the AIG dataset in Fig.\,\ref{fig:data}\,(a) while the latter are from the DHH dataset shown in Fig.\ref{fig:data}\,(b).

At the last row in Fig.\,\ref{fig:Results_compiled}, it shows the line profiles at the center mass of the biggest added tissue at Row\,3 and the corresponding correlation coefficient (CC) comparing to the ground truth. 
Fig.\,\ref{fig:Results_compiled}\,-\,(i,j) is used to refer to the figure in the i$^\text{th}$ row and the j$^\text{th}$ column.

\begin{figure*}
    \centerline{\includegraphics[width=0.8\textwidth]{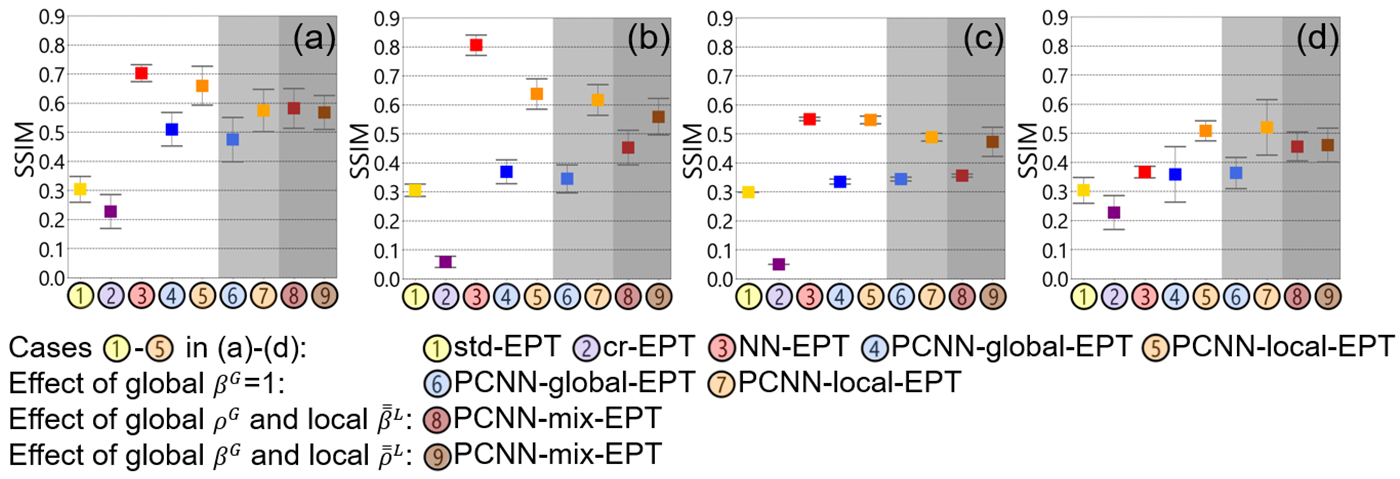}}
    \caption{Means and standard deviations of the SSIM values for the reconstructed conductivity maps in (a) Experiment\,1 with the artificial irregular geometries dataset, (b) Experiment\,1 with the digital human head samples, (c) Experiment\,2, and (d) Experiment\,3. PCNN-local-EPT maintains its accuracy as test samples divert from training samples.}
\label{fig:Results-noiseless}
\end{figure*}

Fig.\,\ref{fig:Results-noiseless}\,(a)-(d) show the means and standard deviations of the reconstruction accuracy for the four cases shown at Row\,1-4 in Fig.\,\ref{fig:Results_compiled}, respectively, when different EPT methods were used, \mycircle{1}\,\,std-EPT, \mycircle{2}\,\,cr-EPT, \mycircle{3}\,\,NN-EPT, \mycircle{4}\,\,PCNN-global-EPT, \mycircle{5}\,\,PCNN-local-EPT. 
Meanwhile, Case-\,\mycircle{6}\,\,and\,\,\mycircle{7} show the situations when only $\boldsymbol{\rho}$ is learned and $\boldsymbol{\beta}$ is set to one, for PCNN-global-EPT ($\mathit{\rho}^\textit{G}$) and PCNN-local-EPT ($\Bar{\Bar{\mathbf{\rho}}}^\textbf{L}$), respectively.
Moreover, the effect of learning a mix of $\mathit{\rho}^\textit{G}$ with $\Bar{\Bar{\mathbf{\beta}}}^\textbf{L}$ is shown in Case-\,\mycircle{8}\,, while the effect of learning a mix of $\mathit{\beta}^\textit{G}$ with $\Bar{\Bar{\mathbf{\rho}}}^\textbf{L}$ is shown in Case-\,\mycircle{9}\,.
Following is a more detailed analysis of the reconstruction results in each experiment. 
\subsubsection{Exp.\,1: 5-fold cross validation}
\label{subsubsec:kfold}
As comparison, Fig.\,\ref{fig:Results_compiled}\,-\,(1,1-6), and Fig.\,\ref{fig:Results_compiled}\,-\,(2,1-6) show the reconstruction results of test sample \#\,26 from the AIG dataset and for test sample \#\,33 from the DHH dataset, respectively. 
The conductivity maps of Column\,4-6, reconstructed by the learning methods show significantly suppressed artifacts, compared with those of Column\,2-3, reconstructed by the non-learning methods. 
Moreover, comparing Column\,5, and 6, it is clear that PCNN-global-EPT has lower reconstruction accuracy than PCNN-local-EPT. When PCNN-local-EPT is further compared with the physics-unaware NN-EPT, for the AIG sample shown at Row\,1, it shows higher accuracy than NN-EPT in Fig.\,\ref{fig:Results_compiled}\,-\,(1,4).
While for DHH samples that have low geometrical variation, PCNN-local-EPT (Fig.\,\ref{fig:Results_compiled}\,-\,(2,6)) results in a lower contrast than NN-EPT (Fig.\,\ref{fig:Results_compiled}\,-\,(2,4)). 
The line profiles of $\Bar{\Bar{\mathbf{\rho}}}^\textbf{L}$ and $\Bar{\Bar{\mathbf{\beta}}}^\textbf{L}$ in Column\,10 and 12, respectively, show that variations corresponding to the boundaries.

Fig.\,\ref{fig:Results-noiseless}\,(a) and (b) show the SSIM values for the 5-fold cross-validation experiment for the AIG and the DHH dataset, respectively.
In Fig.\,\ref{fig:Results-noiseless}\,(a) and (b), comparing the data among \mycircle{1}\,-\,\mycircle{5}\,, the learning methods have higher accuracy than the non-learning methods.
As shown in Fig.\,\ref{fig:Results-noiseless}\,(a), (b), both PCNN-global-EPT and PCNN-local-EPT shows a lower accuracy when the samples are changed from the more diverse AIG dataset in Fig.\,\ref{fig:data}\,(a) to the less diverse DHH dataset in Fig.\,\ref{fig:data}\,(b). 

Moreover, in both Fig.\ref{fig:Results-noiseless}\,(a) and (b), among the learning methods, NN-EPT shows the highest SSIM, because, in Exp.\,1, the test samples do not deviate from the training samples.
In either Fig.\,\ref{fig:Results-noiseless}\,(a) or (b), the comparison between the results of Group \mycircle{6}\,-\,\mycircle{7} and Group \mycircle{4}\,-\,\mycircle{5}\, shows that when only $\boldsymbol{\rho}$ is learned, globally or locally, and $\boldsymbol{\beta}\negmedspace=\negmedspace1$, the accuracy of the reconstructions is hardly compromised.
Furthermore, comparison among the results of Case-\,\mycircle{4}\,\,where both coefficients are globally learned, Case-\,\mycircle{8}\,\,and Case-\,\mycircle{9} where one of the coefficients is globally learned, and Case-\,\mycircle{5}\,\,where both coefficients are locally learned, shows that locally learned coefficients led to higher accuracy, and the accuracy is highest when both coefficients are locally learned. 

\subsubsection{Exp.\,2: Modified test samples}
\label{subsubsec:Pathos}
Fig.\ref{fig:Results_compiled}\,-\,(3,1-6) shows the reconstructed EP of test sample \#\,75.
This sample is selected since it presents both a rotated head model with embedded spherical tissues (in dashed red boxes), making it a very difficult generalization test. 
Among the line profiles under comparison, both analytic methods show rippling artifacts whereas all the learning methods show fewer ripples.
Among the learning methods, the PCNN-global-EPT line profile in Fig.\ref{fig:Results_compiled}\,-\,(5,5) shows low contrast in the reconstruction, unable to produce the contrast of the cerebrospinal fluid (CSF) tissue with a conductivity of 2.14 S$/$m.
The NN-EPT's line profile in Fig.\,\ref{fig:Results_compiled}\,-\,(5,4) shows an over-estimation of the contrast near the beginning of the line profile and an unstable profile for the added mass shown by a slope in the mass's reconstructed conductivity.
The results by the proposed PCNN-local-EPT method in Fig.\ref{fig:Results_compiled}\,-\,(3,6) and (5,6) shows a slight overestimation of the CSF contrast, and an apt contrast of the added mass. 
Moreover, the red arrows on the reconstructed conductivity and generated $\Bar{\Bar{\rho}}^\textbf{L}$ profile in Fig.\ref{fig:Results_compiled}\,-\,(5,6), and (3,10) show how $\Bar{\Bar{\mathbf{\rho}}}^\textbf{L}$ adapts to the pathological mass by increasing its value near the boundary of the added mass to smooth the transition to the surrounding white matter tissue.

As shown in Fig.\,\ref{fig:Results-noiseless}\,(c) at \mycircle{1}\,-\,\mycircle{5}\,, trend similar to those in Fig.\,\ref{fig:Results-noiseless}\,(b), with the major difference that both NN-EPT and PCNN-local-EPT show an accuracy reduction. 
Due to the significant accuracy drop of NN-EPT, its accuracy matches that of the proposed PCNN-local-EPT.
For the effect of $\boldsymbol{\beta}\negmedspace=\negmedspace1$, when $\mathit{\rho}^\textit{G}$ was used, as shown in a comparison between Case\,-\,\mycircle{6}\, and \mycircle{4}\,, the difference of their SSIM is negligible. 
On the other hand, when $\Bar{\Bar{\mathbf{\rho}}}^\textbf{L}$ is used, comparison between Case\,-\,\mycircle{7} and \mycircle{5}\, shows that there is a slight drop of SSIM, reflecting the importance of both coefficients working together for EPs reconstruction. 
Furthermore, comparison between the results of  Group\,-\,\mycircle{8}\,-\,\mycircle{9} and Group\,-\,\mycircle{4}\,-\,\mycircle{5}\, shows the importance of $\Bar{\Bar{\mathbf{\rho}}}^\textbf{L}$ to increase the accuracy of the reconstruction.

\subsubsection{Exp.\,3: Dataset generalization}
\label{subsubsec: X-over}
Among the three generalization tests, this test is the most challenging one, since the test samples deviate the most from the training samples, which requires more general rules from the learning system for EPs reconstruction.
Fig.\,\ref{fig:Results_compiled}\,-\,(4,1-6) shows the reconstructed conductivity maps of sample \#\,26.
This is the same sample as that used for the 5-fold cross-validation task shown at Fig.\,\ref{fig:Results_compiled}\,-\,(1,1-6). A direct comparison between the figures in these two rows can be made to show the effect of the similarities between test and training samples on the reconstructions. 
Comparing Fig.\,\ref{fig:Results_compiled}\,-\,(4,3) and Fig.\,\ref{fig:Results_compiled}\,-\,(4,5), the reconstruction of PCNN-global-EPT presents similar artifacts to that of cr-EPT because of the low value of the generated global diffusion coefficient ($\mathit{\rho}^\textit{G}\negmedspace\approx\negmedspace0$).
In Fig.\ref{fig:Results_compiled}\,-\,(4,6), the proposed PCNN-local-EPT shows a reconstruction that has comparable accuracy to that in Exp.\,1 as shown in Fig.\ref{fig:Results_compiled}\,-\,(1,6). 
In contrast, the reconstructed conductivity maps using NN-EPT in Fig.\ref{fig:Results_compiled}\,-\,(4,4), shows prominent unexpected artifacts which do not appear in Exp.\,1.

Fig.\,\ref{fig:Results-noiseless}\,(d) shows the accuracy for all test samples in Exp.\,3. 
Among Case\,-\,\mycircle{3}\,-\,\mycircle{5}\,, it can be seen that the accuracy of the NN-EPT reconstruction decreases further, and it becomes comparable to the analytic methods. Moreover, it becomes lower compared to both of the proposed PCNN-EPT. 
For the effect of setting the convection coefficient to one ($\boldsymbol{\beta}\negmedspace=\negmedspace1$) with a learned $\mathit{\rho}^\textit{G}$ in \mycircle{6}\,, by comparing to that in which $\mathit{\rho}^\textit{G}$ is learned in \mycircle{4}\,, it shows a similar accuracy.
Furthermore, the accuracy is similar in the cases where the diffusion coefficient is locally learned in PCNN-local-EPT in \mycircle{5} and \mycircle{7}\,.
Comparing the data from \mycircle{8}\,-\,\mycircle{9} to \mycircle{4}\,-\,\mycircle{5}\,, Case\,-\,\mycircle{5} when both coefficients are locally learned shows the highest SSIM and Case\,-\,\mycircle{4} when both coefficients are globally learned shows the lowest SSIM, which indicates the importance of both coefficients acting in conjunction locally to improve the accuracy.
\begin{figure}[!t]
	\centerline{\includegraphics[width=0.5\textwidth]{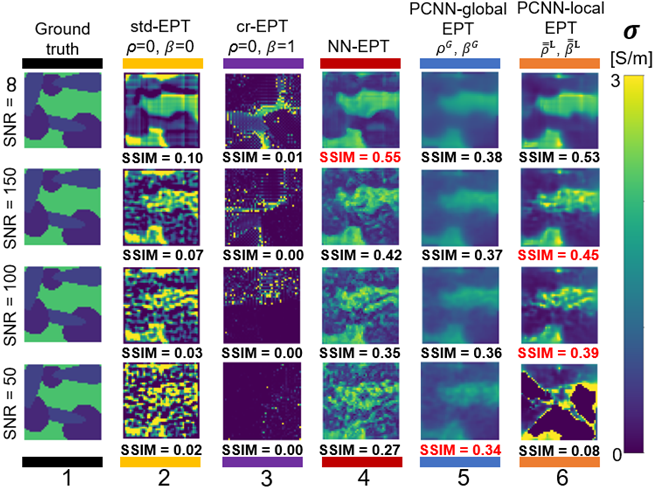}}
	\caption{Reconstructed EPS of Sample\,\#\,10 at SNR levels of $\infty$, 150, 100 and 50 by using different MREPT methods. SSIM values are shown for each reconstruction. PCNN-local-EPT shows highest SSIM at 100 SNR, a common noise level in clinical MRI.}
	\label{fig:Results_noise2}
\end{figure}
\begin{figure*}[t]
    \centerline{\includegraphics[width=\textwidth]{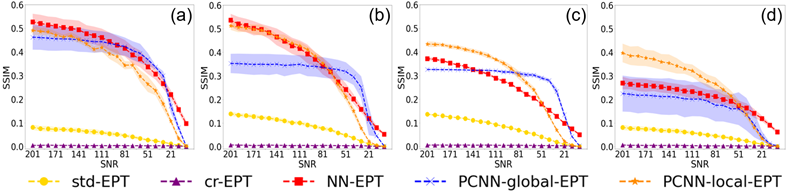}}
     \caption{Mean and standard deviation SSIM values for all methods under comparison, (a) Experiment\,1 with the artificial irregular geometries dataset, (b) Experiment\,1 with the digital human head dataset, (c) Experiment\,2, and (d) Experiment\,3. The test samples are reconstructed at a decreased SNR from 201 to 1 with a step of 10. PCNN-local-EPT retains its noise robustness as test samples divert from training samples.}
\label{fig:Results-noise}
\end{figure*}
\subsection{Examination of Noise Robustness}
\label{subsubsec:Noise robustness}
Fig.\,\ref{fig:Results_noise2} shows the reconstructed EPs of sample \#\,10 from the AIG dataset in Exp.\,1, at four SNR levels (SNR\,=\,$\infty, 150, 100, 50$) orderly at the rows. 
The ground truth conductivity is shown in Column\,1, followed by the reconstructions by non-learning methods (std-EPT and cr-EPT) in Column\,2, and 3, respectively. Column\,4 shows the physics-unaware NN-EPT reconstruction, and Columns\,5, and 6 show the proposed PCNN-global-EPT and PCNN-local-EPT, respectively.
std-EPT reconstruction in Column\,2 shows its characteristic boundary artifacts even without noise. As the noise increases, spurious artifacts appear related to the numerical derivation of the Laplacian term of the formulation in \eqref{eqn:ph-std-EPT}. 
cr-EPT reconstruction is shown in Column\,3, in this case, numerical artifacts are apparent without noise and as the noise increases, the reconstruction is filled by artifacts. 
NN-EPT reconstruction in Column\,4 shows the accuracy of the NN-EPT methods without noise. However, as noise increases, the reconstructions show random artifacts.
PCNN-global-EPT reconstructions in Column\,5 shows low contrast in the reconstructions at different SNRs, which shows the advantage of being robust across the noise spectrum.
The PCNN-local-EPT reconstructions in Column\,6 show a higher contrast without noise. Meanwhile, as the SNR decreases to 100 which is the standard noise level of most MRI scanners, the SSIM of the reconstructed conductivity is kept at 0.39, which is the highest across all the methods. It is noted that when the SNR decreases to 50, a destructive artifact appears. 

The means and standard deviations of the SSIM of the EPs in Exp.\,1-3 from 201 to 1 SNRs are compiled in Fig.\ref{fig:Results-noise}. Fig.\ref{fig:Results-noise}\,(a) and (b) shows the data for Exp.\,1. 
Whereas Fig.\ref{fig:Results-noise}\,(c) and (d) show the data for Exp.\,2 and 3, respectively.
Fig.\ref{fig:Results-noise}\,(a) shows that the learning methods over-perform the non-learning methods across the different noise levels. 
When a less diverse dataset was used, as shown in Fig.\ref{fig:Results-noise}\,(b), the difference between the learning and the non-learning methods is maintained. However, the accuracy of PCNN-global-EPT is lower than that of the other data-driven models when the SNR is higher than 70. Nevertheless, it shows a very small decrease in accuracy as the SNR decreases, therefore, it outperforms the other two learning methods when the SNR is decreased to below 70. 
Comparing results in Fig.\ref{fig:Results-noise}\,(b) to Fig.\ref{fig:Results-noise}\,(c), the SSIM of both the NN-EPT and PCNN-local-EPT drop. However, one highlight is that the proposed PCNN-local-EPT shows higher SSIM compared to NN-EPT up to an SNR of 50.
When the test samples are highly different from the training samples, as shown in Fig.\ref{fig:Results-noise}\,(d) PCNN-local-EPT maintains a similar trend as in Exp.\,2, which can be interpreted as the robustness of the proposed method to the diversity between the test and training samples even in noise conditions.

\section{Discussion}
\label{sec:discussion}
In this section, the effects of trained $\boldsymbol\rho$ and $\boldsymbol\beta$ on the quality of reconstructed EPs, the generalization, and noise robustness of the proposed method are further discussed. 

\subsection{Diffusion Coefficient ($\boldsymbol\rho$) \& Convection Coefficient ($\boldsymbol\beta$)}
\label{stab-coeff}
In the proposed PCNN-EPT, the coefficients $\boldsymbol\rho$ and $\boldsymbol\beta$ regulate the diffusion term ($\nabla^2\gamma$) and one part of the convection term ($\nabla\phi^{\text{tr}}\nabla\gamma$), respectively.

The $\boldsymbol\rho$ coefficient determines how one pixel affects its neighborhood. When pixel-wise $\boldsymbol\rho$ (local $\Bar{\Bar{\mathbf{\rho}}}^\textbf{L}$) is applied, the viscosity effect can be set up depending on the situation of each region for a good reconstruction of EPs. Specifically, at the boundaries, it can be set to high values to avoid boundary-related artifacts, whereas, at homogeneous regions, it can have low values to mitigate small numerical artifacts. This is well reflected by Fig.\,\ref{fig:Results_compiled}\,-\,(1,10) and (2,10), the line profiles of normalized local diffusion coefficient, $\Bar{\Bar{\mathbf{\rho}}}^\textbf{L}$, learned for two samples (one from the AIG dataset and the other from the DHH dataset) in Exp.\,1.

On top of a local $\boldsymbol\rho$, when $\boldsymbol\beta$ is learned locally and simultaneously in conjunction with $\boldsymbol\rho$, effective mutual counteraction of $\boldsymbol\rho$ and $\boldsymbol\beta$ are attained, which can be disclosed by comparing the line profiles of learned $\Bar{\Bar{\mathbf{\beta}}}^\textbf{L}$ (Fig.\,\ref{fig:Results_compiled}\,-\,(1,12) and (2,12)) and those of learned $\Bar{\Bar{\mathbf{\rho}}}^\textbf{L}$ (Fig.\,\ref{fig:Results_compiled}\,-\,(1,10) and (2,10)).
This can be further understood through the comparison of different cases in Fig.\,\ref{fig:Results-noiseless}\,(a)\,-\,(b) for Exp.\,1. Case-\,\mycircle{5}\,, in which both coefficients are locally learned, shows higher reconstruction accuracy than those in Case-\,\mycircle{7}\,,\,-\,\mycircle{9}\,, and -\,\mycircle{4} in which either one of the coefficients is globally set or learned. This is true for Exp.\,2, (see Fig.\,\ref{fig:Results-noiseless}\,(c)), when test samples contain additional tissues unseen in training samples. 
However, in Exp.\,3, when test samples deviate largely from the training ones, this advantage does not exist. As shown in Fig.\,\ref{fig:Results-noiseless}\,(d), the difference in accuracy among Case-\,\mycircle{5}\,,\,-\,\mycircle{7} and -\,\mycircle{9} becomes closer. This is owing to that, a fixed $\boldsymbol\beta$ setup can reduce the possibilities of impairing reconstruction with erroneous coefficients generated, especially for samples very different from the training ones. 

\subsection{Generalization}
\label{gen}
The proposed PCNN-EPT involves the formulation originated from std-EPT in \eqref{eqn:ph-std-EPT} and cr-EPT in \eqref{eqn:ph-cr-EPT}, which have severe boundary artifacts due to assumptions in the derivation process and their inflexible representation (fixed coefficients), as shown in the reconstructed EPs at Column\,2, and 3 in Fig.\,\ref{fig:Results_compiled}. By introducing the two multi-dimensional variable coefficients that are required to be optimized with data, considerable flexibility is achieved, which can be made clear by comparing Column\,5, and 6 with Column\,2, and 3 in Fig.\,\ref{fig:Results_compiled}.

More importantly, the coupling of the optimization of the two multi-dimensional variable coefficients with the analytic model offers PCNN-EPT generality, which is hard to obtain for end-to-end learning and critical to the future clinical application of this method.
As shown in the results of Exp.\,2 (Fig.\,\ref{fig:Results_compiled}\,-\,(3,1-6) and (5,1-6)), for the additional small tissues (illustrated with red squares in Ground Truth in Fig.\,\ref{fig:Results_compiled}\,-\,(3,1)) unseen in the training samples, analytic models (Column\,2 and 3 of Fig.\,\ref{fig:Results_compiled} for std-EPT and cr-EPT, respectively) could reconstruct roughly the additional tissues, however, big errors occurred around them, as indicated by the low correlation coefficients of their line profiles across one of the additional tissues. 
These errors were effectively handled (as shown in Fig.\,\ref{fig:Results_compiled}\,-\,(3,6) and (5,6)), with the help of the $\Bar{\Bar{\mathbf{\rho}}}^\textbf{L}$ and $\Bar{\Bar{\mathbf{\beta}}}^\textbf{L}$ optimized by the NNs in the PCNN-EPT, as presented by the line profile (Fig.\,\ref{fig:Results_compiled}\,-\,(3,10) and (3,12)) of the normalized $\Bar{\Bar{\mathbf{\rho}}}^\textbf{L}$ and $\Bar{\Bar{\mathbf{\beta}}}^\textbf{L}$. 
Therefore, it is reasonable to claim that the coupling of physics-based formulation and the learning for multi-dimensional variable coefficients plays the most important role for high generality. 
This has been proved again by the results of Exp.\,3, shown in Fig.\,\ref{fig:Results_compiled}\,-\,(4,1-6), in which the PCNN-local-EPT presents the highest SSIM value. Although, the line profile of the $\Bar{\Bar{\mathbf{\rho}}}^\textbf{L}$, Fig.\ref{fig:Results_compiled}\,-\,(4,10), shows smaller values around boundaries than those in Fig.\ref{fig:Results_compiled}\,-\,(1,10), it plays a role to mitigate the artifacts, as reflected by Fig.\ref{fig:Results_compiled}\,-\,(4,6), compared with Fig.\ref{fig:Results_compiled}\,-\,(4,2-5).
\begin{figure*}[!h]
	\centering
	\includegraphics[width=0.8\textwidth]{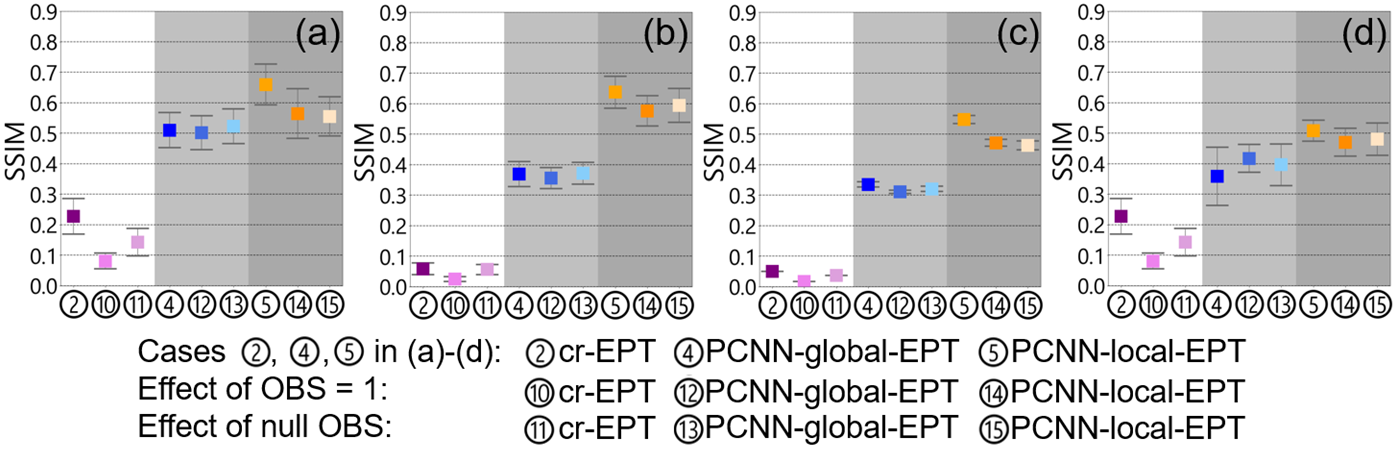}
	\caption{Means and standard deviations of the SSIM values for the reconstructed conductivity maps in (a) Experiment\,1 with the artificial irregular geometries dataset, (b) Experiment\,1 with the digital human head samples, (c) Experiment\,2, and (d) Experiment\,3 at various outer boundary settings (OBS) conditions. Accurate OBS, increase the accuracy of the model slightly, however, is shown that even without OBS the reconstruction is accurate.}
	\label{fig:Results-OBS}
\end{figure*}

\subsection{Noise Robustness}
It is noted that the noise robustness of PCNN-local-EPT is comparable to that of NN-EPT in Exp.\,1 (Fig.\ref{fig:Results-noise}\,(a), (b) and Fig.\,\ref{fig:Results_noise2}\, Column\,4, 5, 6), but higher until SNR of test samples decreases to 51 in Exp.\,2 and 3 (Fig.\ref{fig:Results-noise}\,(c), (d)). 

NN-EPT generates more spurious noise-related artifacts as the noise level increases, as seen at Column\,4 of Fig.\,\ref{fig:Results_noise2}.
This is because NN-EPT maps its input to the conductivity image, based on pattern matching rules acquired by the NN, and thus noise is transferred directly to the reconstructed image in an interpolated manner, and further, it leads to low generalization.
The problem might be exacerbated by multiple sources of noise during the process to obtain $B_1^+$ for MREPT\,\cite{Hampe2020}.

On the other hand, for the proposed PCNN-EPT, when the test samples are noisy, even though the noise may affect both the stabilization coefficients and terms in the analytic model, it can finally be suppressed by the stabilization coefficients.
That is why, in Exp. 2 and 3, even if the $\Bar{\Bar{\mathbf{\rho}}}^\textbf{L}$ and $\Bar{\Bar{\mathbf{\beta}}}^\textbf{L}$ could not be accurately recalled because the test samples contain structures partially or completely different from those in the training samples, PCNN-EPT could achieve comparably accurate reconstruction.

In summary, the proposed method not only inherits the noise robustness of stabilized cr-EPT formulation\,\cite{Li2017}, implemented by the global stabilization coefficients (Column\,5 of Fig.\,\ref{fig:Results_noise2}), but also, greatly improves it by proposing and developing methods to optimize the local stabilization coefficients.

\section{Conclusion}
\label{conclusion}
In this paper, we propose a physics-aware EPT method, PCNN-EPT (i.e., physics-coupled neural network electrical property tomography) that effectively couples the stabilized cr-EPT method with two coefficient generating NNs for optimizing the two cooperating functioning coefficients, $\boldsymbol{\rho}$ and $\boldsymbol{\beta}$, with the most flexibility to produce accurate reconstructions. 
It has been shown that this proposed approach can achieve high reconstruction accuracy for two simulated datasets, even in the difficult case that the test samples deviate the most from the training samples and in a noisy environment, which is challenging for an end-to-end approach. It is the first time that these two coefficients are optimized together, especially locally, taking into consideration different needs for tissue boundaries. As a result, the proposed method shows promising generality and noise robustness.  
Furthermore, while pioneering data-driven optimization coupled to analytic models to make the EPs reconstruction explainable, noise-robust, and generalizable, our approach also provides insights to the understanding of ideal data-dependent coefficients of analytic models.
Moving forward, further research is needed to investigate its suitability to clinical data in various measurement environments.

\appendices
\section{Implementation details}
\label{subsec:Implementation details}

\subsection{Outer boundary setting}
\label{subsubsec:Outer boundary setting}
Furthermore, to solve the cr-EPT formulation a set of linear equations in the $\boldsymbol{Ax}\negmedspace=\negmedspace\boldsymbol{b}$ form are solved to generate a predicted conductivity map.
To solve this set of linear equations, outer boundary settings (OBS) can be applied to improve the reconstruction accuracy. 
Previous studies\,\cite{Ider2020} have argued that the effect of the OBS is trivial because the matrix is well-conditioned.
The effect of OBS for cr-EPT, and the proposed PCNN-EPT both in its global and local presentation was investigated. 
In Fig.\,\ref{fig:Results-OBS}, the cases when the OBS was null, set to 1 [S/m], or taken from the ground truth conductivity are shown for all experiments for the methods where the OBS can be utilized (cr-EPT, PCNN-global-EPT, and PCNN-local-EPT).
In Fig.\,\ref{fig:Results-OBS}\,(a),(d), the cr-EPT reconstruction in Cases-\,\mycircle{2}\,,\,\,\mycircle{10}\,\,and\,\,\mycircle{11} for the artificial geometries dataset shows very big variations in accuracy, while for the digital head models in Fig.\,\ref{fig:Results-OBS}\,(b),(c) the accuracy is very low irrespective of the conditions, mainly to this samples having more boundaries that increase the chances of producing artifacts.
Moreover, the PCNN-global-EPT accuracy in Fig.\,\ref{fig:Results-OBS}\,(a)-(d) for Cases-\,\mycircle{4}\,,\,\,\mycircle{12}\,\,and\,\,\mycircle{13} shows similar accuracy for all cases because most of them produce extremely low contrast reconstruction irregardless of the OBS. 
Finally, the PCNN-local-EPT accuracy in Fig.\,\ref{fig:Results-OBS}\,(a)-(d) for Cases-\,\mycircle{5}\,,\,\,\mycircle{14}\,\,and\,\,\mycircle{15}\,, there is a slight accuracy decay from when the ground truth conductivity is used for the OBS to the other conditions for all experiments. 
\begin{figure*}[!h]
	\centering
	\includegraphics[width=0.8\textwidth]{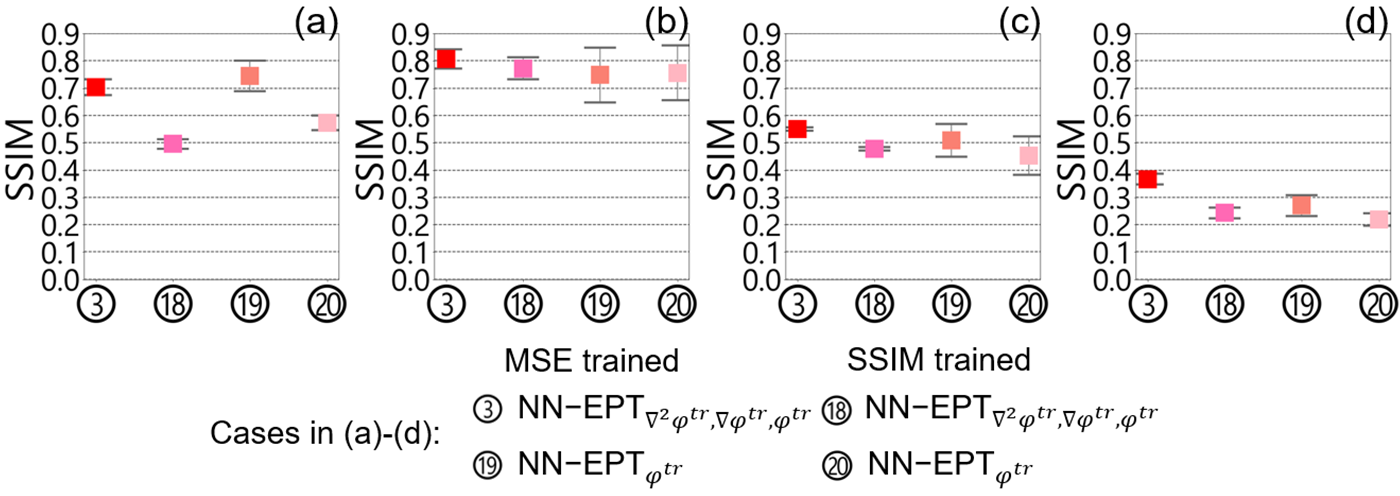}
	\caption{Means and standard deviations of the SSIM values for the reconstructed conductivity maps in (a) Experiment\,1 with the artificial irregular geometries dataset, (b) Experiment\,1 with the digital human head samples, (c) Experiment\,2, and (d) Experiment\,3 for the black-box NN-EPT for MSE and SSIM as loss function during training and $\phi_{\text{tr}}$ and $\nabla^2\phi^{tr}$, $\nabla\phi^{tr}$, and $\phi^{tr}$ as inputs. It is shown that when MSE and $\nabla^2\phi^{tr}$, $\nabla\phi^{tr}$, and $\phi^{tr}$ as input is used for NN-EPT, the reconstruction accuracy increases.}
	\label{fig:Results-unet2}
\end{figure*}

\subsection{Training details}
\label{subsubsec:Training details}
Next, to produce the backpropagation of the error, PyTorch$^\copyright$\,\cite{Paszke2019} is used. This NN framework creates computational graphs that engrave operations in a hierarchical chart to enable automatic differentiation. 
The NNs are trained for a maximum of 1000 epochs and the best accuracy model according to the train samples is saved for testing. We use a learning rate annealing technique, in which, if the loss stops decreasing for 50 epochs the learning rate is halved. Moreover, to accelerate the training, in the proposed PCNN-EPT, the learning rate is also halved if the loss increases due to an accumulation of artifacts on the reconstruction. This further prevents getting stuck at local optima.
The training time ranges from half an hour for the NN-EPT to one hour for the proposed PCNN-EPT in either global or local forms, all training was produced on an Nvidia GeForce RTX 2070 GPU.

\subsection{U-net architecture}
\label{subsubsec:U-net architecture}
The "U-net"\,\cite{Ronneberger2015} is composed of a mirroring compression and dilation set of filters. 
First, the inputs go through a double convolution applying one 5x5 padded convolution layer with a hyperbolic tangent activation function. Consequently, a 3x3 padded convolution layer with a leaky Relu activation is applied. 
Then, the same double convolution set-up is applied to double the number of channels, plus a 2x2 max pooling down-sampling operation to increase the perceptive field of the convolution operation. This process is repeated four times. 
Next, the outputs of the compression filters are up-sampled. We up-sample by a doubling size bilinear interpolation. 
Then, the up-sampled feature is concatenated with the matching down-sampled feature from the compression filters and finally, the concatenated input goes through another double convolution set-up as described above. 
This process is also repeated four times to match the output size.
Lastly, the up-sampled output goes through a 1x1 convolution to compress the number of channels and produce the output. 
The values of the weights are {[64, 128, 256, 512, 256, 128, 64]} respectively.

\section{NN-EPT}
\label{subsec:NN-EPT}
To produce a point of independent comparison against an end-to-end NN-EPT model, where one NN outputs the final conductivity map directly, without any intermediate step is investigated.
Two types of inputs were investigated. 
One model received the transceive phase as input (NN-EPT$_{\phi^{tr}}$), to avoid the noise explosion associated with the numerical derivative calculation. 
The second model, besides the transceive phase, adds the transceive phase derivatives (NN-EPT$_{\nabla^2\phi^{tr},\nabla\phi^{tr},\phi^{tr}}$), concatenated to produce a multi-channel input, since based on many analytic methods these features bring relevant information. 
Furthermore, the NN-EPT can be trained with different loss functions, we compared training with the usual loss function for end-to-end methods MSE, and SSIM since it was the one selected for the proposed method.
Because all models are representative of the end-to-end approach, we used the one with the best accuracy in the generalization tasks from Experiment\,2 and Experiment\,3 for a point of comparison with the proposed method.
The compiled results for all experiments are shown in Fig.\,\ref{fig:Results-unet2}. 
In it we can see that the NN-EPT methods have slight differences in each experiments, although, the NN-EPT$_{\nabla^2\phi^{tr},\nabla\phi^{tr},\phi^{tr}}$ trained with SSIM in Case-\,\mycircle{3}\,, over-performs for the generalization tasks in Fig.\,\ref{fig:Results-unet2}\,(c) and (d) compared to Case-\,\mycircle{19}\, when the same input is trained with SSIM or to Cases-\,\mycircle{18} and \mycircle{19}, when the $\phi^{tr}$ is used as input with MSE or SSIM, respectively. 

\section*{Acknowledgment}
The authors would like to thank Ph.D. Stefano Mandija for his constructive discussion about numerical simulations and dataset construction.

\bibliographystyle{IEEEtran}
\bibliography{signal}

\begin{IEEEbiography}[{\includegraphics[width=1in,height=1.25in,clip,keepaspectratio]{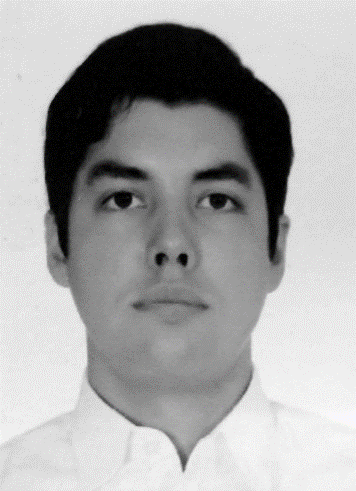}}] {Adan Jafet Garcia Inda} was born in Guadalajara, Jalisco, Mexico in 1991. He received the B.Eng. in biomedical engineering from the University of Guadalajara, Jalisco, Mexico in 2009, received a M.Eng. in medical systems engineering from Chiba University, Chiba, Japan in 2019. He is currently pursuing the Ph.D. degree in Medical Engineering from Chiba University, Chiba, Japan. His research interest include artificial intelligence, machine learning, quantitative magnetic resonance imaging, and point of care healthcare technologies. \end{IEEEbiography}

\begin{IEEEbiography}[{\includegraphics[width=1in,height=1.25in,clip,keepaspectratio]{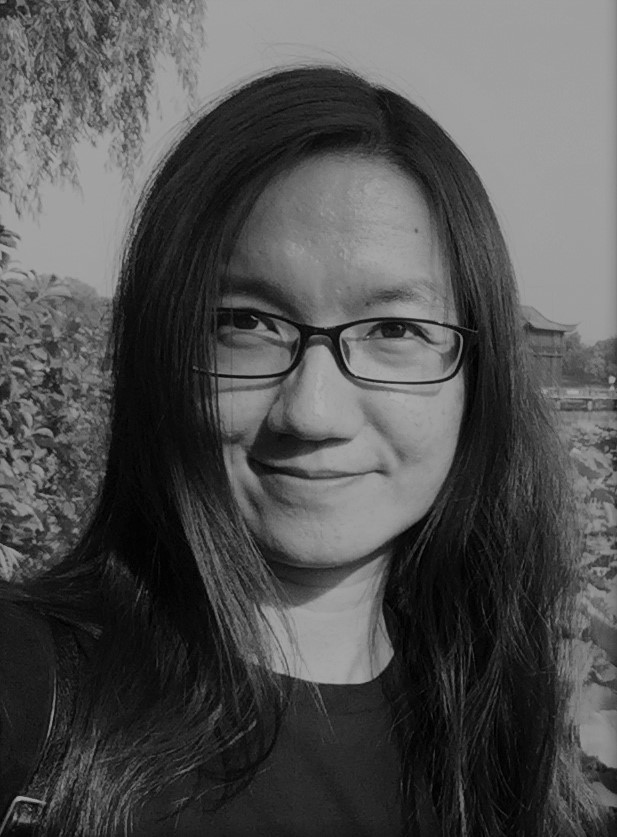}}] {Shao Ying Huang} received the B.Eng., M. Eng., and Ph.D. degree from Nanyang Technological University, Singapore in 2003, 2006, and 2011, respectively. She joined the University of Hong Kong in 2010 working on computational electromagnetics (EM), and Massachusetts Institute of Technology in 2012 working on magnetic resonance imaging (MRI) related EM problems, both as a postdoctoral fellow. She is an associate professor in the pillar of Engineering Product Development, Singapore University of Technology and Design. She is an adjunct assistant professor in the department of surgery in National University of Singapore, Singapore. Her research interests include radiofrequency(RF)/microwave noninvasive/contactless sensing, low-field MRI, non-linear MRI image reconstructions, and RF aspects of MRI, MR electrical property tomography, wireless power transfer, wideband RF/microwave components. \end{IEEEbiography}

\begin{IEEEbiography}[{\includegraphics[width=1in,height=1.25in,clip,keepaspectratio]{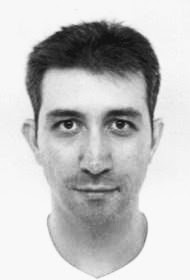}}] {Nevrez \.{I}mamo\u{g}lu} is currently employed as a Researcher at National Institute of Advanced Industrial Science and Technology (AIST), Tokyo, Japan since April 2016. Before AIST, he was with RIKEN Brain Science Institute as a Researcher \& JSPS Foreign Postdoctoral Fellow in 2015-2016. He received the Ph.D. degree from Chiba University (2015), Japan, with a Japanese Government scholarship (MEXT). He was also with School of Computer Eng., Nanyang Technbological University, Singapore as Research Associate 2010-2011. He received the M.S. and B.S. degrees from TOBB University of Economics and Technology (Ankara, Turkey) and Cankaya University (Ankara, Turkey) in 2010 and 2007. His research interests include applications of computer vision and machine learning, recently with special interests on deep learning for multi-modal or multi-temporal signal/image analysis, visual attention (saliency detection), classification/object detection/segmentation on various sensory data (3D, optical, multispectral, hyperspectral). \end{IEEEbiography}

\begin{IEEEbiography}[{\includegraphics[width=1in,height=1.25in,clip,keepaspectratio]{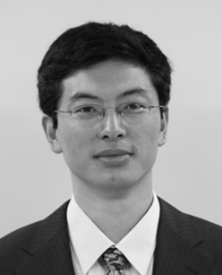}}] {Wenwei Yu} received B.Eng. and M.Eng. degrees from Shanghai Jiao Tong University in 1989 and 1992, respectively, a Ph.D. degree in system information engineering from Hokkaido University, Japan, in 1997, and a Ph.D. degree in Rehabilitation Medical Science, Hokkaido University, Japan, in 2003. He served as an Assistant Professor with the System Information Engineering Department, School of Engineering, Hokkaido University, from 1999 to 2003. He was an Exchange Research Fellow with the Center for Neuroscience, University of Alberta, Canada, in 2003, supported by the Researcher Exchange Program, Japanese Society for Promotion of Science (JSPS). He has been an Associate Professor with the Department of Medical System Engineering, School of Engineering, Chiba University, Japan, since 2004 and has also been a Professor since 2009. Since 2006, he has been with the AI Lab, Zurich University, Switzerland, as a Visiting Professor, supported by the Japanese Society for Promotion of Science (JSPS). He has authored and coauthored more than 150 papers in refereed journals, book chapters, and more than 170 international conference papers. His research interests include neuroprosthetics, rehabilitation robotics, motor control, and biomedical signal processing. He is a member of the Robot Society of Japan (RSJ) and the Japanese Society for Medical and Biological Engineering (JSMB). \end{IEEEbiography}

\end{document}